\RequirePackage{fix-cm}
\documentclass[twocolumn,epjc3]{svjour3}  
\smartqed  % flush right qed marks, e.g. at end of proof
\usepackage{mathtools}
\usepackage{xcolor}
\usepackage{booktabs}
\usepackage{tabularx} 
\usepackage[bookmarks=false,hypertexnames=true]{hyperref} 
\usepackage{tikz}
\usepackage{blindtext}

\usepackage{placeins}
\usepackage{subfig}
\usepackage{lineno}
\RequirePackage{graphicx}

\usepackage{afterpage}

\newcommand{\jpsi}{$J/\psi$}

\newcommand{\s}[1]{$\sqrt{s} = #1 $}
\newcommand{\kStar}{k^{*}}
\RequirePackage{mathptmx}      % use Times fonts if available on your TeX system
\begin{document}

%\title{Extraction of inelastic cross section of a quarkonium with hadrons from quarkonium-hadron femtoscopic function measurement}
\title{A technique to study the elastic and inelastic interaction of quarkonium with hadrons using femtoscopic correlations}

%\subtitle{Do you have a subtitle?\\ If so, write it here}

%\titlerunning{Short form of title}        % if too long for running head

\author{Marzieh Bahmani\thanksref{e1,addr1}
        \and
        Daniel Kiko\l{}a \thanksref{e2,addr1} 
        \and
        Leszek Kosarzewski \thanksref{e3,addr3}        %etc.
}

%\thankstext{t1}{Grants or other notes
%about the article that should go on the front page should be
%placed here. General acknowledgments should be placed at the end of the article.
\thankstext{e1}{e-mail: marzieh.bahmani@pw.edu.pl}
\thankstext{e2}{e-mail: Daniel.Kikola@pw.edu.pl}
\thankstext{e3}{e-mail: kosarles@fjfi.cvut.cz}
%\authorrunning{Short form of author list} % if too long for running head

\institute{Warsaw University of Technology, Faculty of Physics, Koszykowa 75, Warsaw, Poland \label{addr1}
           \and
           Czech Technical University in Prague, Faculty of Nuclear Sciences and Physical Engineering, Brehova 7, Prague, Czech Republic \label{addr3}         
           %\emph{Present Address:} if needed\label{addr3}
}

\date{Received: date / Accepted: date}
% The correct dates will be entered by the editor

\maketitle

\begin{abstract}
We present a method for the measurement of parameters of elastic and inelastic interactions of charmonium with hadrons. 
In this technique, we use femtoscopic analysis of charmonium-hadron correlations at low relative momentum and the Lednicky-Lyuboshitz analytical model to extract the interaction parameters. We argue that such a study is already feasible in the LHCb experiment at the LHC, and we discuss the prospects for studies in the STAR at RHIC and other experiments at the LHC.

\keywords{charmonium \and femtoscopic correlations \and \jpsi-hadron interaction \and  Lednicky-Lyuboshitz model}
% \PACS{PACS code1 \and PACS code2 \and more}
% \subclass{MSC code1 \and MSC code2 \and more}
\end{abstract}

\section{Introduction}

\label{intro}

The main purpose of experimental high-energy nuclear physics is to investigate the properties of the Quark-Gluon Plasma (QGP). In the ``normal'' matter, constituents of the visible Universe (partons, i.e., quarks and gluons) are confined within hadrons. In contrast, the QGP is a matter in local thermal equilibrium with quark and gluon degrees of freedom. The partons, not hadrons, define the properties of the QGP. Such a state of matter existed in the early Universe, microseconds after the Big Bang and we can create it for a short while in heavy-ion collisions with high enough energy density.
Experiments at the SPS (Super Proton Synchrotron), RHIC (Relativistic Heavy Ion Collider), and the LHC (Large Hadron Collider) demonstrated that the QGP has unique properties~\cite{Busza:2018rrf}. Among others, it behaves like (almost) perfect and the most vortical fluid known so far~\cite{STAR:2017ckg}. 

The study of partonic matter in a laboratory is a difficult task, since a small droplet of the QGP matter produced in high-energy nuclear collisions only exists for a short time (of the order of 10 fm). Then it cools down and subsequently partons form hadrons which are registered by experiment. Thus, the effects of hadronic phase always accompany the QGP signals.

During the past 25 years, physicists developed a variety of approaches to access the properties of the QGP, for example,  studies of the modification of the energy spectra of jets and heavy quarks in heavy-ion collisions in comparison with proton-proton interactions, or analysis of the momentum anisotropy of particles in the final state of the collisions. Here we focus on using charmonium and bottomonium states (quarkonium for short) as a probe of the QGP, and more precisely, on ``calibration'' of such a probe.

The idea of using the production of a \jpsi\ meson (a bound state of $c$ and $\overline{c}$ quarks) to study the properties of the QGP was proposed by Matsui and Satz~\cite{Matsui}. They demonstrated that the binding potential of $c$ and $\overline{c}$ quarks will be screened in the partonic matter, which would cause a suppression of \jpsi ~meson production (per nucleon-nucleon collision) in heavy-ion reactions with respect to yield in nucleon-nucleon interactions. The same argument applies to other members of the charmonium and bottomonium families~\cite{FaccioliP}, however the suppression of a given state depends on the energy density (hence temperature) of the partonic matter. Thus, simultaneous measurement of the production of \jpsi, $\psi(2s)$, $\varUpsilon(1S), \varUpsilon(2S),\varUpsilon(3S)$ and other quarkonium states could provide information about thermodynamic properties of the QGP within this paradigm.

This idea of using quarkonium to probe the QGP properties became popular because of its elegance, but the reality of heavy-ion collisions is more complicated. Firstly, there are other possible ways of charmonium and bottomonium interaction with the partonic matter (see the review~\cite{sapore} and the references therein) and there are dynamical effects that one needs to take into account. Specifically, the production rates of $c\overline{c}$ and $b\overline{b}$ pairs depend on the parton distribution in a heavy nucleus (which is an initial-state effect) and a bound state ($c\overline{c}$ or $b\overline{b}$) could be destroyed by passing through the ``cold'' matter of a nucleus present in a collision (this effect is called nuclear absorption). Finally, charmonium or bottomonium could be destroyed in the last, hadronic phase of the reaction. 

The production of charmonium and bottomonium states has been extensively studied experimentally in the last\sloppy decades (see for example ~\cite{Adare:2006ns,STAR1,Chatrchyan:2012lxa,Khachatryan:2016xxp,Adam:2016rdg,Acharya:2019lkh,Abelev:2013yxa,Chatrchyan:2013nza,Adare:2010fn,Aaij:2017cqq}, to quantify both effects from the hot partonic and ``cold'' matter.  A plethora of data is available; yet, no theoretical model can describe all results. One source of the problem is the entanglement of processes that affect quarkonium production, which need to be included in such a model. Specifically, the data collected both at the  RHIC~\cite{Adare:2013ezl} and at the LHC~\cite{Chatrchyan:2013nza} energies suggest that the interaction of quarkonium with hadrons in the final state is an important factor and it requires attention. We propose a novel experimental approach that provides information about the elastic and inelastic (destructive) interactions of charmonium and bottomonium with hadrons in the final stage of the collision. With the results of such study, we will be one step closer to understanding the quarkonium interaction within the QGP.  

We propose to study the correlations at low relative momentum (so-called femtoscopic correlations) of quarkonium with hadrons in proton-proton collisions. The correlations are sensitive to space-time properties of the particle emission region and interactions in the final state; and the existing formalism allows for the calculation of the interaction parameters (see for example~\cite{Adamczyk:2015hza}). Thus, one can \textit{measure}  the cross section for elastic and inelastic interactions of quarkonium in the hadronic phase. We focus on the \jpsi-hadron case because \jpsi ~is produced copiously in high-energy collisions and its measurement is straightforward, but the reasoning applies to other quarkonium states.   

In this paper, in  section \ref{sec:jpsi:theory}, we briefly review models for $J/\psi$-hadron interactions to present the context of our study. Then in  section \ref{sec:theory}, we discuss the theoretical basics of the femtoscopic correlations. We introduce the \sloppy Lednicky-Lyuboshitz analytical model that links an experimental correlation function to final-state strong interaction parameters, and we outline how to calculate the elastic and inelastic cross section from these parameters. In  section ~\ref{sec:fis}, we perform a feasibility study for the measurement of the  \jpsi-hadron femtoscopic function and evaluation of the cross section for \jpsi-hadron interactions for two experiments: STAR at RHIC and LHCb at the LHC, using the data they already collected. The effect of the feed-down from higher charmonium states, non-femtoscopic background and the results are also presented in this section. Then, we discuss  the prospects for such measurements in the near future in  section \ref{sec:prospects}. Finally, we conclude the paper in  section \ref{sec:con}.

\section{Models of \jpsi ~interaction with hadrons in high energy nuclear collisions}
\label{sec:jpsi:theory}

We outlined in the previous section that the interaction of charmonium with hadrons in the final state of nuclear collisions (specifically heavy-ion collisions) is an important factor in the interpretation of experiment studies of the QGP. The elastic scattering can change the momentum distribution of \jpsi, thus it mimics the energy loss in the nuclear matter, while the inelastic (destructive) interaction leads to a suppression of the observed yields, thus it resembles the destruction of the quarkonium in the partonic phase. 

Those final-state interactions are extremely difficult to quantify because they are convoluted with other effects.
In the standard approach, the yields in  proton-nucleus (p+A) collisions where the effect occurs are compared to a baseline measurement in proton-proton reactions, and then  the results should be checked for compatibility with the phenomenological model (see for example~\cite{phenix_abs,lansberg_abs,ramona_abs}). In the p+A collisions, there are already initial state effects and interactions with nuclear matter, which obscure the final observations. 

The comover interaction model~\cite{ferreiro} is an example of a theoretical calculation, where quarkonium-hadron interactions play an important role. In this model, the probability of inelastic quarkonium-hadron interactions increases as the hadron density increases. In this model, interaction with the medium is quantified by a cross section $\sigma_{\mathrm{abs}}^{\mathrm{comover}}$ for the breakup of \jpsi ~with comoving matter (regardless whether it consists of partons or hadrons). In general, the $\sigma_{\mathrm{abs}}$ is an external parameter, which has to be fixed, either in comparison with data or in theoretical calculations. 

There is a handful of theoretical calculations of the $\sigma_{\mathrm{abs}}$ parameter, and they vary significantly. For the extended discussion of the models for quarkonium dissociation in hadronic matter, we redirect the reader to the review~\cite{Rapp:2008tf}, here we provide a few examples. A calculation based on the meson exchange model in a chiral Lagrangian calculation~\cite{meson_exchg,Lin:1999ad} yields the values of absorption cross section $0.8-3\:\mathrm{mb}$ for \jpsi-$\pi$ interaction and $0.1-1\:\mathrm{mb}$ for \jpsi-$\rho$. Another approach based on the extended Nambu Jona-Lasinio model~\cite{njl} yield values of $0.1-1\:\mathrm{mb}$ for \jpsi-$\pi$ absorption. 

Those values of $\sigma_{\mathrm{abs}}$ are \textit{assumed} to cause insignificant effects in heavy ion reactions~\cite{Rapp:2008tf}, although these results and the corresponding conjecture were never verified experimentally. Moreover, the cross section for the interaction of higher charmonium states with hadrons is expected to be much larger~\cite{Du:2015wha}, which is in agreement with the data~\cite{Adare:2013ezl}.   

In this paper, we present a method for a direct measurement of these cross sections using quarkonium-hadron femtoscopy. Such a study will  provide a good opportunity to test the calculations of charmonium-hadron interactions and improve models for quarkonium production in heavy-ion collisions.

\section{\jpsi-hadron femtoscopy and formalism for extracting the interaction parameters}
\label{sec:theory}

To measure the space-time characteristics of the particle-emitting region in hadron collisions, the approach of using the correlation of two identical bosons was introduced almost six decades ago~\cite{Goldhaber}. Since then, the method of momentum correlation was refined in the seminal works of Kopylov and Podgoretsky (~\cite{Kopylov:1972qw,Kopylov:1975,Kopylov:1974th}), and extended to non-identical particle pairs (please see \cite{Lisa} for a comprehensive review). The correlations at low relative momentum are also sensitive to Coulomb and strong interactions between particles in a pair~\cite{Lednicky:1981su,Lednicky:2008zz}, thus they provide rich information about collisions.

The femtoscopic correlation function for two particles in general is defined as:
\begin{equation}
C({\bf p_{a}},{\bf p_{b}}) = \frac{P_{2}({\bf p_{a}},{\bf p_{b}})}{P_{1}({\bf p_{a}})P_1({\bf p_{b}})}
\label{eq:ckstar}
\end{equation}

where $P_{1}$ is the probability of observing a particle with a given momentum ${\bf p_{a}}$ and $P_{2}$ is the conditional probability of a particle with momentum ${\bf p_{b}}$ being observed if a particle of momentum ${\bf p_{a}}$ is also observed.

According to~\cite{Lisa} the Eq.~\ref{eq:ckstar} can be rewritten as
\begin{equation}
C(\kStar) = \int d^{3}r^{*} S(r^{*})|\Psi{(\vec{r^{*}},\kStar)}|^{2}
\label{eq:ckstar2}
\end{equation}
where $\kStar$ is the momentum of a particle in the
pair center-of-mass (c.m.) system, and $S(r^{*})$ is the source function. The $S(r^{*})$ is a distribution of
the relative distance $r^{*}$ of particles in the pair c.m. system and it contains all space-time information about the emission source. The relative wave function of the particle pair beyond the range of the potential is denoted by $\Psi{(\vec{r^{*}},\kStar)}$ has the form
\begin{equation}
\label{form5}
 \Psi (\vec{r^{*}},\kStar) \doteq e^{i\vec{k^{*}} \cdot \vec{r^{*}}}+ \frac{f^{S}(k^{*})}{r^{*}} e^{-ik^{*} \cdot r^{*}},
\end{equation}
which represents the stationary solution of the scattering
problem.

The correlation function is considered as
a square of the wave function $\Psi$  averaged over  the  relative distance vector
${r}^*$ of the emitters in the pair c.m. system  and over the pair total spin. It is the asymptotic form of a superposition
of the plane and outgoing spherical waves.

The form of $\Psi (\vec{r^{*}},\kStar)$ is estimated by taking into account the outside range of the strong interaction potential,  and the s-wave part of the scattered wave.This allows the internal structure of the strong interaction potential to be neglected and be assumed that it is spherically symmetric. In this case, only the magnitude of the potential is relevant.
Then s-wave scattering amplitude  in the effective range approximation ~\cite{effR} at considerably
small $k^*$ values can be written as:
\begin{eqnarray}
\label{fs}
f^{S}(k^{*}) = ( \frac{1}{f^{S}_{0}}
+ \frac{1}{2} d^{S}_{0} k^{*2} - i k^{*} )^{-1},
\end{eqnarray}
where $f^{S}_{0}$ is the scattering length and
$d^{S}_{0}$ is the effective radius for a given total spin of 1 (triplet state) or 0 (singlet state). With the assumption that
particles are emitted unpolarized, the fraction $\rho$ of pairs with a given spin are $\rho_0=1/4$ for pairs in the singlet state and
$\rho_1=3/4$ for the triplet state.
%RL

The Lednicky-Lyuboshitz (L-L) analytical model~\cite{Lednicky:1981su} connects the correlation function with final-state strong interaction parameters. It assumes $\vec{r^{*}}$
has Gaussian distribution as follows:

\begin{eqnarray}
d^{3}N/d^{3}r^{*} \sim e^{- \vec{ r^{*} }^{2} /4r_{0}^{2}},
\end{eqnarray}
where $r_{0}$ can be considered as the effective radius of the source,
then the correlation function can be calculated analytically:
\begin{equation}
\begin{aligned}
\label{cft}
& C(k^*) = 1 + \sum_S \rho_S\left[\frac12\left|\frac{f^S(k^*)}{r_0}\right|^2\left(1-\frac{d_0^S}{2\sqrt\pi r_0}\right)\right. + \\
& \left.\frac{2\operatorname{Re}(f^S)(\kStar)}{\sqrt\pi r_0}F_1(Qr_0)-\frac{\operatorname{Im} (f^S(k^*))}{r_0}F_2(Qr_0)\right] ,
\end{aligned}
\end{equation}
where
$F_1(z) = \int_0^z dx e^{x^2 - z^2}/z$ and $F_2(z) = (1-e^{-z^2})/z$.
From Eq.~\ref{cft} one can see that this model relates the observed two-particle correlation function to the source size and the s-wave scattering amplitude.

\subsection{Cross section of elastic interactions with the presence of inelastic interactions}

Both elastic and inelastic cross sections for \jpsi-hadron interaction can be calculated from the scattering amplitudes. The scattering amplitudes can in turn be extracted from the experimental data by fitting the \jpsi-hadron femtoscopic correlation function with the formula in Eq.~\ref{cft}. In the partial wave expansion, the elastic scattering cross section $\sigma_{el}$~\cite{QM} can be calculated with:

\begin{equation}
\begin{aligned}
\label{eq1}
	\sigma_{el} = 4\pi \sum_{l}(2l+1)|f_{l}|^2 \stackrel{l=0}{=} 4\pi |f^{S}(k^{*})|^2
\end{aligned}
\end{equation}

Where $f_{l}$ are the scattering amplitudes for each partial wave $l$, while $f^{S}(\kStar)$ is the s-wave scattering amplitude from Eq.~\ref{fs} in the s-wave approximation ($l=0$).

If a partial wave is absorbed, which means inelastic scattering is present, then the corresponding S-matrix element $S_{l}$ satisfies the  $|S_{l}|^{2}<1$ relation. This matrix element can be expressed in terms of $f_{l}$.

\begin{equation}
\begin{aligned}
\label{eq2}
	S_{l} = 1+2i\kStar f_{l}
\end{aligned}
\end{equation}

Then the Eq.~\ref{eq2} can be plugged into the formula for inelastic scattering cross section $\sigma_{inel}$~\cite{QM}:

\begin{equation}
\begin{aligned}
\label{eq3}
& \sigma_{inel} = \frac{\pi}{k^{*2}} \sum_{l}(2l+1)(1-|S_{l}|^2) = \\
%& \frac{\pi}{k^{*2}} \sum_{l}(2l+1)(1-1+4\kStar\operatorname{Im}(f_{l})-4k^{*2}|f_{l}|^{2}) = \\
& 4\pi \sum_{l}(2l+1) \left( \frac{\operatorname{Im}(f_{l})}{\kStar}-|f_{l}|^{2} \right) 
\end{aligned}
\end{equation}

Again, in the s-wave approximation, this gives:

\begin{equation}
\begin{aligned}
\label{eq4}
 \sigma_{inel} \stackrel{l=0}{=} 4\pi \left( \frac{\operatorname{Im}(f^{S})}{\kStar}-|f^{S}(\kStar)|^{2} \right)
\end{aligned}
\end{equation}

Finally, the total scattering cross section is a sum of both elastic and inelastic cross sections:

\begin{equation}
\begin{aligned}
\label{eq5}
	\sigma_{tot} = \sigma_{inel}+\sigma_{el} = \frac{4\pi}{\kStar}\operatorname{Im}(f^{S}(\kStar))
\end{aligned}
\end{equation}

It has to be noted that in the case of singlet and triplet states, each cross section component is a sum of cross sections weighted by the spin fraction $\rho_{i}$ for the corresponding state as:

\begin{equation}
\begin{aligned}
\label{eq6}
	\sigma = \sum_{i}\rho_{i}\sigma_{i}
\end{aligned}
\end{equation}

\begin{figure}[hbt!]
  \includegraphics[scale=0.9]{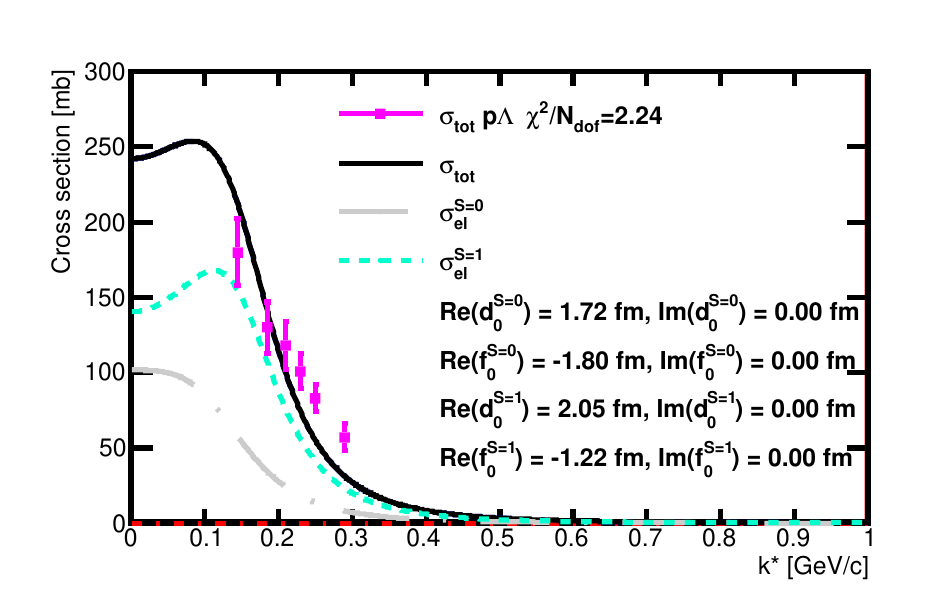}
\caption{ Cross sections vs. $\kStar$ calculated in the s-wave approximation using parameters from ~\cite{juelich_eft}, compared to the experimental data ~\cite{pLambda_data} for $p-\Lambda$ interactions. }
\label{fig:xsec:1}       % Give a unique label
\end{figure}

We tested these formulas for the parameters obtained in the case of $p-\Lambda$ interactions in the Effective Field Theory approach~\cite{juelich_eft} as shown in Fig.~\ref{fig:xsec:1}. The calculations describe the data reasonably well, however there are small discrepancies as $\kStar$ increases. These discrepancies are due to the limitations of the s-wave approximation which is valid mostly at low-$\kStar$.

It is worth to mention that the requirement on the matrix element $|S_{l}|^{2}<1$ for inelastic scattering leads to a constrain on the imaginary part of the scattering length. The $\operatorname{Im} f_{0}$ must be positive.

\begin{figure*}[hbt!]
\includegraphics[scale=0.9]{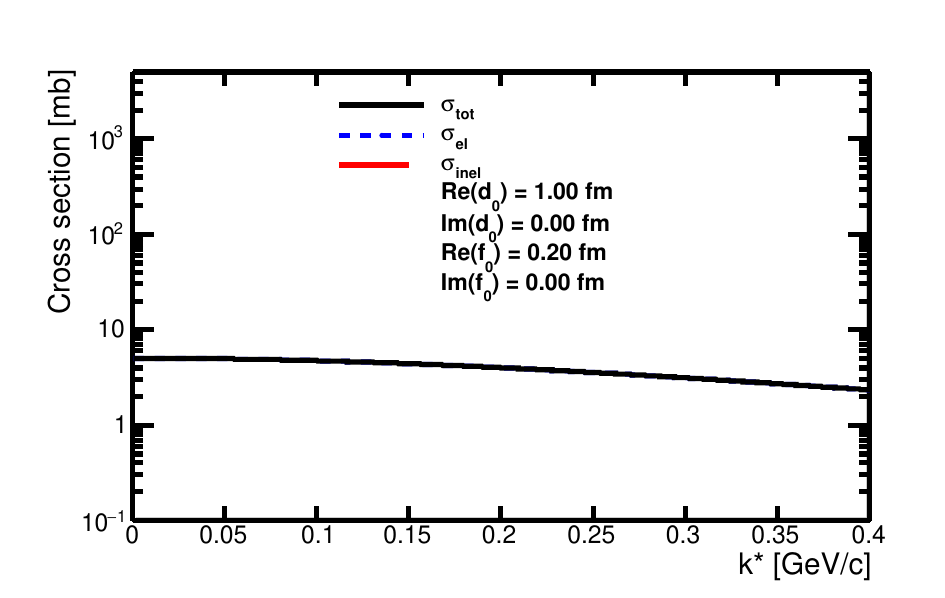}	
\includegraphics[scale=0.9]{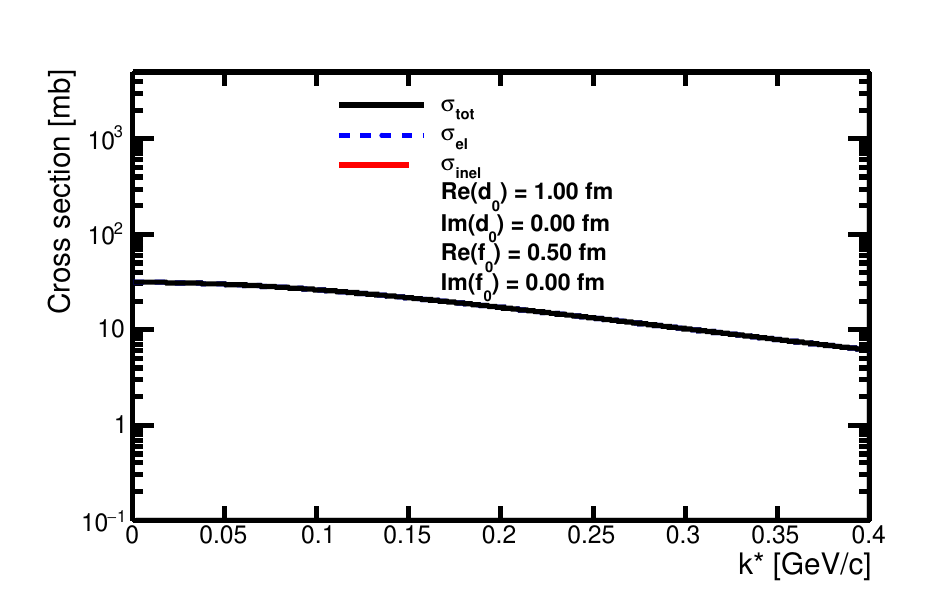}
\includegraphics[scale=0.9]{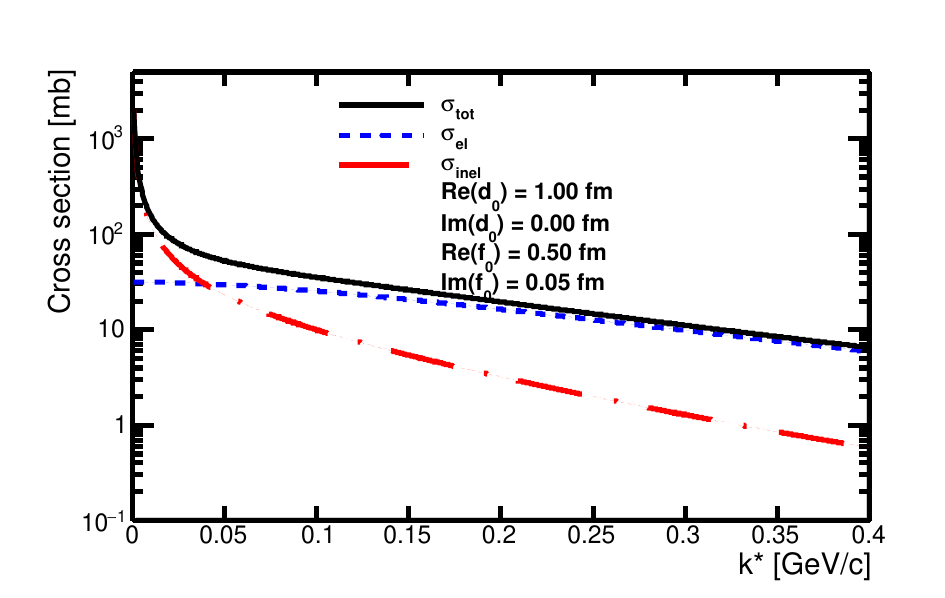}
\includegraphics[scale=0.9]{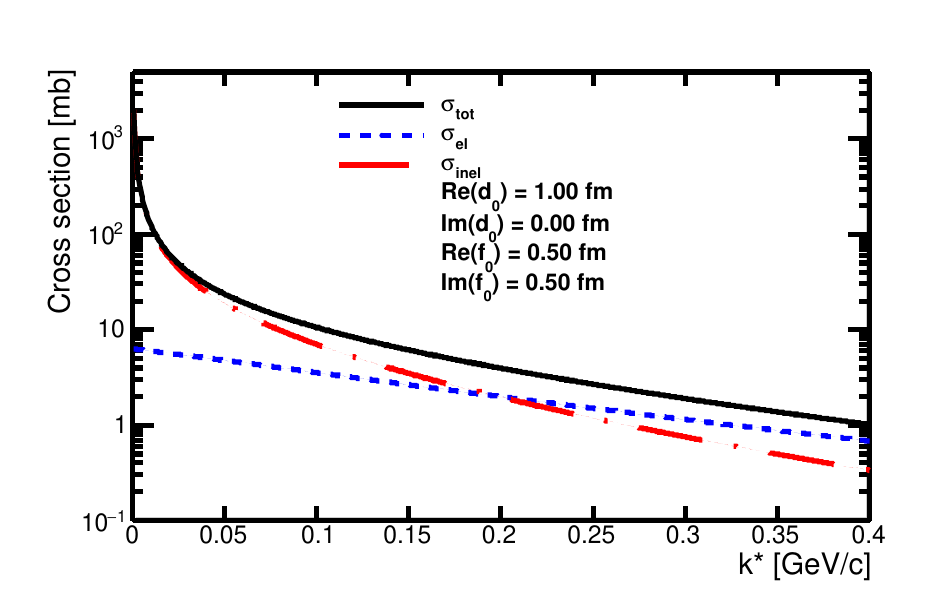}
\caption{The sensitivity of elastic and inelastic cross section for \jpsi-hadron interaction to the parameters of interaction: the real and imaginary part of scattering length.}
\label{fig:xsec:parameter}
\end{figure*}

When $\operatorname{Im} f_{0} = 0\:\mathrm{fm}$, the inelastic cross section vanishes. 
Figure ~\ref{fig:xsec:parameter} shows examples of cross section calculations for different values of $\operatorname{Re} f_{0}$, $\operatorname{Im} f_{0}$, and $\operatorname{Re} d_{0}$.

\section{Feasibility study for \jpsi-hadron femtoscopy at the STAR and the LHCb experiments}
\label{sec:fis}
In this section, we investigate the feasibility of measurements of  \jpsi-hadron femtoscopic correlation function, hence the measurements of the \jpsi\ breakup cross section for the studies by the STAR and the LHCb experiments in $pp$ collisions at the center-of-mass energy of 500 GeV and 8 TeV respectively. We consider that the integrated luminosity available at LHCb is $L_{int}=2082$ $pb^{-1}$ from 2012 data taking period~\cite{LHCb-Facts}, and STAR collected data with $L_{int} = 400$~$pb^{-1}$ in 2017 data~\cite{star_fut}. In addition, we review the case of STAR collecting f $L_{int}= 2.2$ $fb^{-1}$  which is foreseen for 2023 data taking campaign at\s{510}~GeV

\subsection{Simulation setup }
\label{sectionSim}

For this study, the simulation samples were generated with Pythia 8.2 configured with the parameters within the LHCb and  STAR experiments acceptance as following:

\paragraph{LHCb setup:}

The event samples of $pp$ collisions with the energy of 8 TeV, hadron $p_{T} >$ 0.5 GeV, and hadron pseudorapidity 2 $<\eta <$ 5 were considered for the LHCb acceptance. We selected \jpsi\ within the rapidity range of $2 < y^{\mathrm {J}/\psi}< 5$ for the reconstruction via $J/\psi \rightarrow \mu^+\mu^-$ decay channel. We assumed that the hadron and \jpsi\ reconstruction efficiency have approximately constant values of 0.96 and 0.25, respectively, which is inspired by the reported performance of the LHCb experiment~\cite{lhcb_per,Aaij:2012asz}. 

\paragraph{STAR setup:}
The event samples of $pp$ collisions with the energy of \s{500} GeV, hadron $p_{T} >$ 0.2 GeV, and hadron pseudorapidity $|\eta| <$ 1 were considered. We analyzed \jpsi\ produced at mid-rapidity: with $|y^{\mathrm {J}/\psi}|< 0.4$ for $J/\psi \rightarrow \mu^+\mu^-$ and $|y^{\mathrm {J}/\psi}|< 1$ for the $J/\psi \rightarrow e^+e^-$ decay channel~\cite{star_1}. The hadron and \jpsi\ reconstruction efficiency are taken from ~\cite{Abelev:2008ab,star_1} and they are applied as a function of transverse momentum. 

From the simulated samples, the $\kStar$ distribution for the given efficiency and acceptance of each experiment were modeled. These distributions include the non-femtoscopic background. The sources of non-femtoscopic correlations are any resonances that decay to the same particles, which in our case are mostly various B mesons decay to \jpsi-hadron pairs.

\begin{figure}
\includegraphics[trim={1.5cm 0 1cm 0},clip,scale=0.49]{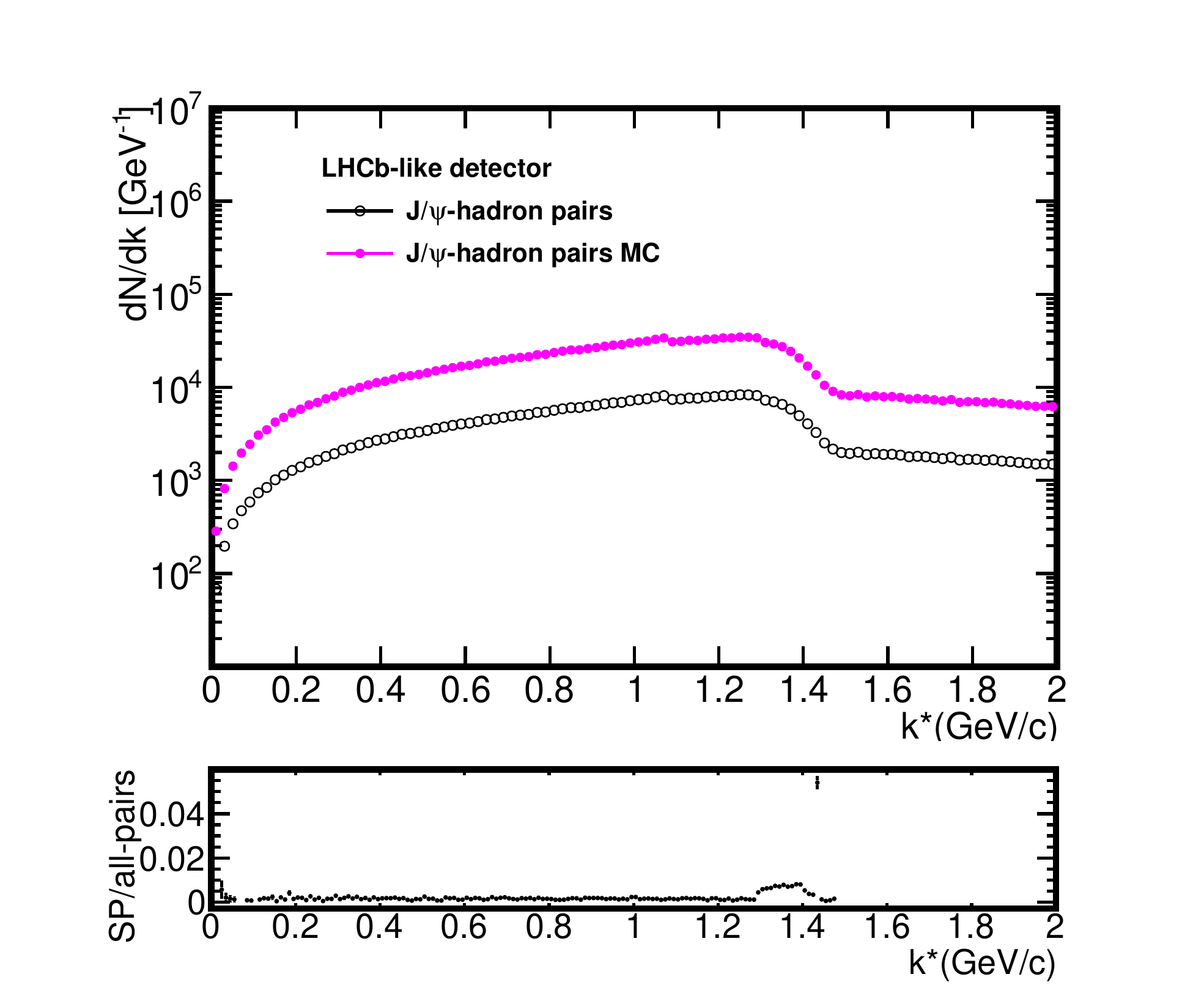}

\caption{Top panel: The distribution of $k^*$ for the LHCb acceptance with and without applying the efficiencies. Bottom panel: the ratio of \jpsi-hadron from the same parent (SP) to all pairs. }
\label{fig:lhcb_kstar}      
\end{figure}

\begin{figure}

  \includegraphics[trim={1.5cm 0 1cm 0},clip,scale=0.49]{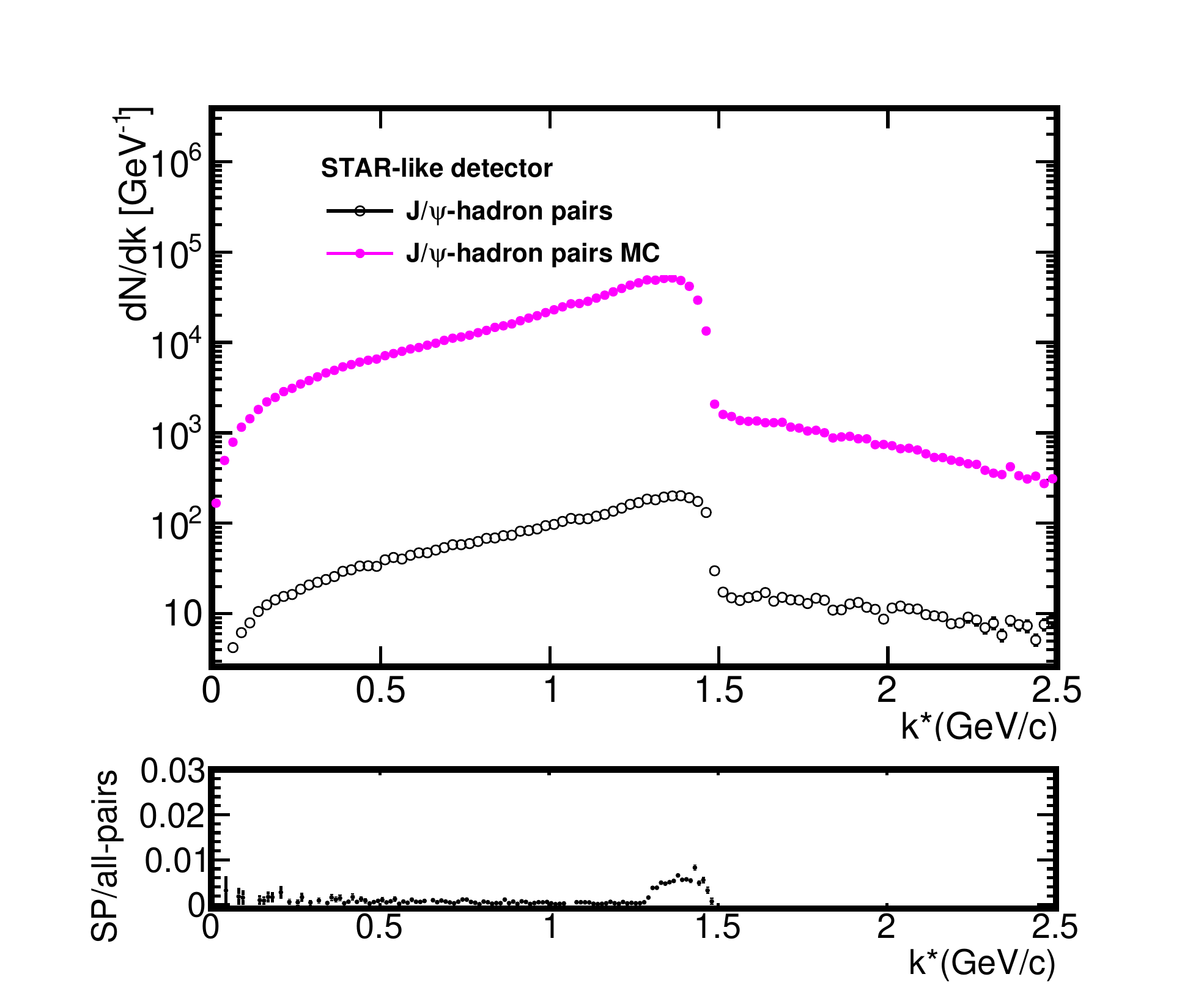}

\caption{Top panel: The distribution of $k^*$ for the STAR acceptance for $J/\psi \rightarrow e^+e^-$ reconstruction with and without applying the efficiencies. Bottom panel: the ratio of \jpsi-hadron from the same parent (SP) to all pairs. }
\label{fig:star_kstar}       
\end{figure}

The $k^*$ distributions modeled for LHCb-like and STAR-like experiments are shown in Fig. \ref{fig:lhcb_kstar} and \ref{fig:star_kstar} respectively,  with and without applying the efficiencies in the top plots. We observed, that there is no significant difference between the shape of $k^*$ distributions for $J/\psi \rightarrow \mu^+\mu^-$ and $J/\psi \rightarrow e^+e^-$ reconstruction at STAR. The bottom panels in Fig. \ref{fig:lhcb_kstar} and \ref{fig:star_kstar} show the ratio of \jpsi-hadron pairs from the same parent to all pairs, which indicates that at low $k^*$ the non-femtoscopic background (\jpsi-hadron pairs from resonance decays) is negligible.\\

We calculated the expected number of \jpsi\ mesons for the feasibility studies by taking the \jpsi\ counts reported by the LHCb and the STAR experiments for $L_{int} = 18.4$ $pb^{-1}$ and $L_{int} = 22.1$ $pb^{-1}$, respectively, and scaling them up to the expected integrated luminosity. 

The number of \jpsi-hadron pairs expected for the femtoscopic measurement is calculated as following: 
\begin{equation}
\langle N_{J/\psi-h}\rangle= N_{J/\psi} \times \langle N_{h}\rangle
\end{equation} 
where $N_{J/\psi}$ and $\langle N_{h}\rangle$ are the \jpsi\ raw yield and the estimated  mean number of charged hadrons observed in the events that contain \jpsi. We used Pythia to calculate these values and we obtained $\langle N_{h}\rangle = 5.31 \pm 0.01$ for LHCb-like experiment and $\langle N_{h}\rangle = 4.83 \pm 0.01$ for the STAR acceptance. In this approach we assumed that at most  one \jpsi\ is observed in an event. Finally, since we use the range $k^* <$ 0.4 GeV, we applied a correction factor to calculate the $\langle N_{J/\psi-h}\rangle$ usable for the femtoscopy. 

Table~\ref{tab:setup} shows the estimated number of \jpsi\ and \jpsi-hadron pairs for each of the considered cases.
These expected number of \jpsi-hadron pairs and $\kStar$ distribution are used for producing the pseudo-experimental femtoscopic correlation function $C(\kStar)$. For STAR, we limited our feasibility study to the $J/\psi \rightarrow e^+e^-$ channel because the \jpsi\ yields are significantly higher than for  $J/\psi \rightarrow \mu^+\mu^-$.

\begin{table*}[hbt!]
\centering
	
\begin{tabular}{lllllllll}
\toprule

$ $ & $ $ & $ $ & \multicolumn{2}{l}{\scriptsize Published raw \jpsi\ yield and $L_{\textrm int}$}& \multicolumn{2}{c}{ \scriptsize Expected raw \jpsi\ yield and  $L_{\textrm int}$} & $ $ & \scriptsize Expected number of pairs\\

\midrule
\scriptsize Detector & \scriptsize Decay channel &\scriptsize $\sqrt{s}$ [TeV]& \scriptsize  \jpsi\ yield & \scriptsize$L_{\textrm int} [pb^{-1}]$& \scriptsize  $L_{\textrm int} [pb^{-1}]$&\scriptsize $N_{J/\psi}$ $\times 10^{6}$&\scriptsize $\langle N_{h} \rangle$ & \scriptsize $\langle N_{J/\psi-h}\rangle$ $\times 10^{6}$ \\

\midrule
\scriptsize LHCb & $J/\psi \rightarrow \mu^+\mu^-$ & \scriptsize 8 &\scriptsize 2.6$\times 10^{6}$ &\scriptsize 18.4&\scriptsize 2082&\scriptsize 294& \scriptsize 5.31&\scriptsize 1562\\
\scriptsize STAR& $J/\psi \rightarrow e^+e^-$ &\scriptsize 0.5&\scriptsize 9581 &\scriptsize 22.1&\scriptsize 400&\scriptsize 0.173 & \scriptsize 4.82&\scriptsize 0.83\\
\scriptsize STAR & $J/\psi \rightarrow e^+e^-$ &\scriptsize 0.51&\scriptsize 9581&\scriptsize 22.1&\scriptsize 2200&\scriptsize 0.95 & \scriptsize 4.82&\scriptsize 4.6\\
\scriptsize STAR& $J/\psi \rightarrow \mu^+\mu^-$ &\scriptsize 0.51&\scriptsize 1154&\scriptsize 22.0&\scriptsize 2200&\scriptsize 0.115 & \scriptsize 4.82&\scriptsize 0.56\\
\bottomrule  
\end{tabular}
\caption{The expected number of \jpsi-hadron pairs  $\langle N_{J/\psi-h} \rangle$ for LHCb-like ~\cite{lhcb_1,LHCb-Facts} and STAR-like ~\cite{star_1,star_fut} experiments for the data samples collected by LHCb in 2012 data taking period; and by STAR in 2017, and the foreseen run in 2023~\cite{star_fut}. To obtain the $N_{J/\psi}$ for each experiment, we considered the measured raw \jpsi\ yield in data set with a given integrated luminosity, and scaled the yields up to the total available integrated luminosity of the corresponding experiment.}\label{tab:setup}
\end{table*}

We model the $C(\kStar)$  according to the L-L analytical model using several sets of parameters listed in Table \ref{tab:input:params}. The choice of given values is motivated by experimental results for femtoscopic measurements~\cite{cor1,cor2,cor3}. 
We varied the effective radius and the scattering length  of the potential within $d_{0}\epsilon[0,1]$ and $f_{0}\epsilon[0,2]$, to assess a precision of determination of these interaction parameters for different amplitudes and shapes of the correlation function.

\begin{table}[hbt!]
	\centering

	\begin{tabular}{cccc}
		\toprule
		Set No.&$\operatorname{Re}(d_{0}^{S})$ [fm]&$\operatorname{Re}(f_{0}^{S})$ [fm]&$\operatorname{Im}(f_{0}^{S})$[fm]\\
		\midrule
		1 & 1.0 & 0.2 & 0.0 \\
		2 & 1.0 & 0.2 & 0.5 \\
		3 & 1.0 & 0.5 & 0.5 \\
		4 & 1.0 & 1.0 & 0.5 \\
		5 & 0.0 & 0.5 & 1.0 \\
		6 & 0.0 & 1.5 & 1.0 \\
		\bottomrule
	\end{tabular}
	\caption{Parameters of the L-L model which were used for generating femtoscopic correlation functions.
	For each parameter set we assume $r_{0}=1.25$~fm and $\operatorname{Im}(d_{0}^{S})=0$.}
	\label{tab:input:params}
\end{table}

To estimate the statistical precision expected from a measurement of interaction parameters via femtoscopy, pseudo-experimental correlation functions were generated for the LHCb-like and STAR-like experiments for a given set of input parameters. We sampled the $\kStar$ distribution at $\kStar < 0.4$ GeV and created a distribution with weights defined by $C(\kStar)$ with a given set of input parameters from Table~\ref{tab:input:params}. The number of samples is given by the expected yield of \jpsi-hadron pairs. Figures  \ref{fig:corr:fun:LHCb} and \ref{fig:corr:fun:STAR} show the pseudo-experimental $C(\kStar)$ with statistical uncertainties for LHCb- and STAR-like experiments.   

Finally, we fit the correlation functions with the L-L model to extract the scattering length and the effective range. This way we evaluate the statistical uncertainties expected for these parameters in a given experiment, and we can assess the feasibility of such a study. Since the fit is challenging, we made a few simplifications. First, we assume that one will not be able to differentiate between singlet and triplet states, so we assume they are the same. Then, in the fit we fixed $\operatorname{Im}(d_0) = 0$. Finally, we constrain $r_0 = 1.25$~fm, which means that the average emission distance is the same as observed in the study of hadron-hadron femtoscopic correlations~\cite{cor1,cor2,cor3}.

\subsection{Effect of the feed-down from higher states of charmonium}

The experimental evidence of charmonium production on the \jpsi, $\psi(2s)$ and $\chi_{c}$ cross section ~\cite{Faccioli} indicates that $\chi_{cJ}$ and $\psi(2s)$ feed-down  fraction to \jpsi  ~production rates by considering all the correlation and uncertainties are $(25.3 \pm 1.8)\%$ and $(7.5 \pm 0.3)\%$ respectively. The analysis~\cite{Faccioli} shows that $(67.2 \pm 1.9)\%$ of the prompt \jpsi ~yield is due to the directly produced meson.  
According to the 2020 Review of Particle Physics ~\cite{pdg}, the $\chi_{cJ}$ to \jpsi ~happens via radiative decay. Although $\psi(2s)$ decays to \jpsi ~+ $\pi^{+}\pi^{-}$ and \jpsi ~+ $\pi^{0}\pi^{0}$ has the fraction around 52$\%$, for \jpsi +$\eta$ around 3$\%$ and for \jpsi +$\pi^{0}$ around 0.0012$\%$, in the studies on the feed-down fraction of $\psi(2s)$ ~\cite{Faccioli} for lower $p_{T}$ spectrum of the mother particle, it has been measured around 7.5$\%$. 

In general, radiative decays complicate the interpretation of the measurements, but do not generate a background in the proposed study. However, $\psi(2s)$ feed-down should be taken into consideration as a possible source of non-femtoscopic background.

\subsection{Non-femtoscopic background}

The source of non-femtoscopic correlations, that is the background in our study, are the resonances which decay to \jpsi ~+ hadron pairs. Such candidates are B-mesons (the branching ratio of decay to \jpsi ~+ anything $(1.094\pm 0.032)\%$) and the excited state of \jpsi ~such as $\psi(2s)$. We use our Pythia simulations to estimate this effect. Figures~\ref{fig:lhcb_kstar} and~\ref{fig:star_kstar} (bottom plots) show that the low $\kStar$ region is not significantly affected by resonance decays. The feed down from B mesons is visible in the high-$\kStar$ range, outside of the region of our interests.

\subsection{Results}
\label{sub:results}

\begin{figure}
\centering
\subfloat[]{\includegraphics[scale=0.48]{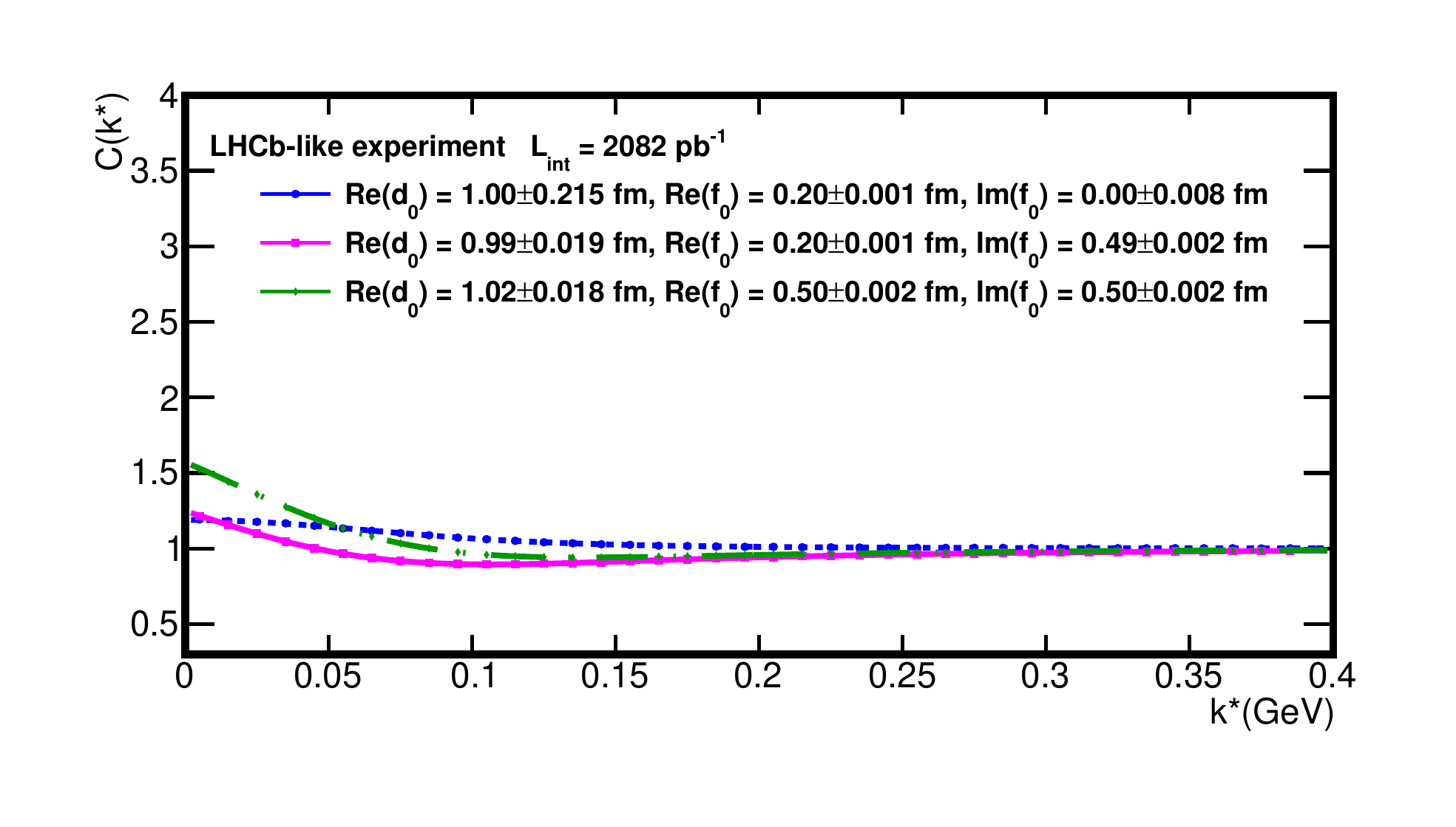}}
\vspace{-0.5cm}
\subfloat[]{\includegraphics[scale=0.48]{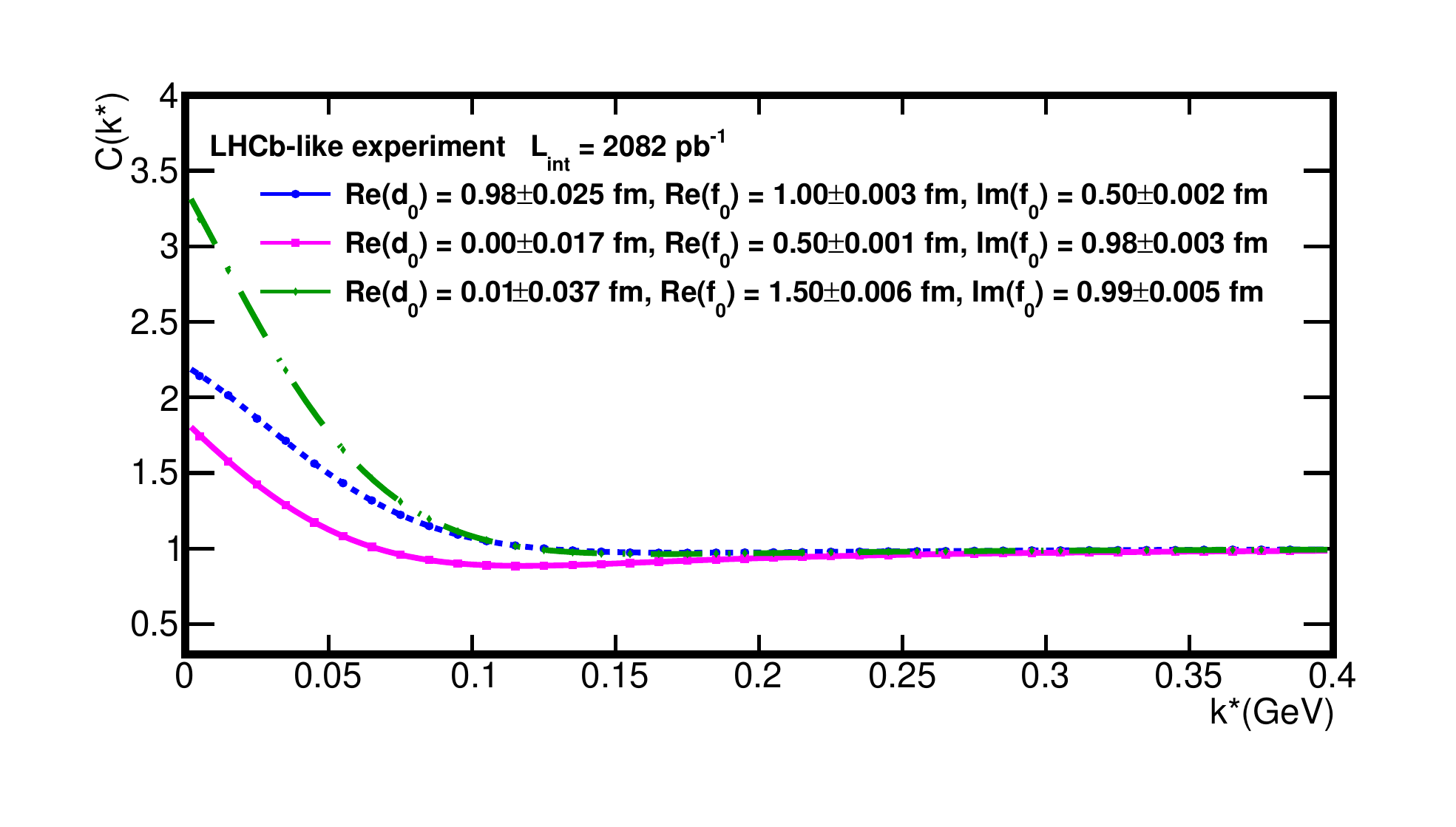}}

\caption{The pseudo-experimental correlation function for the LHCb-like experiment the  plot (a) for the first three sets and  plot (b) for the second three sets  of parameters from Table~\ref{tab:input:params}, and their fit. }
\label{fig:corr:fun:LHCb}       
\end{figure}

\begin{figure}
\centering
 \subfloat[]{ \includegraphics[scale=0.48]{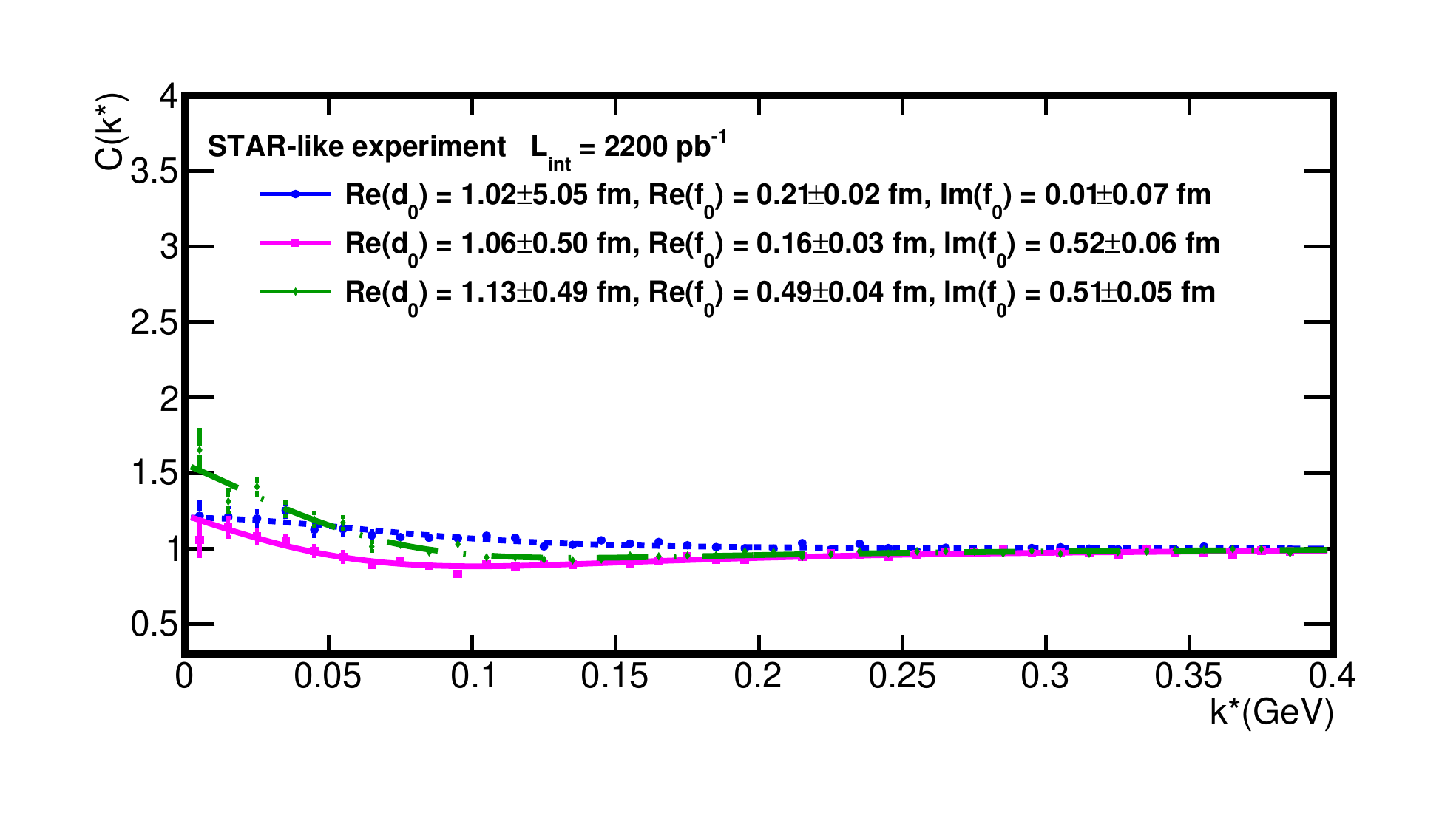}}
 \vspace{-0.5cm}
 \subfloat[]{ \includegraphics[scale=0.48]{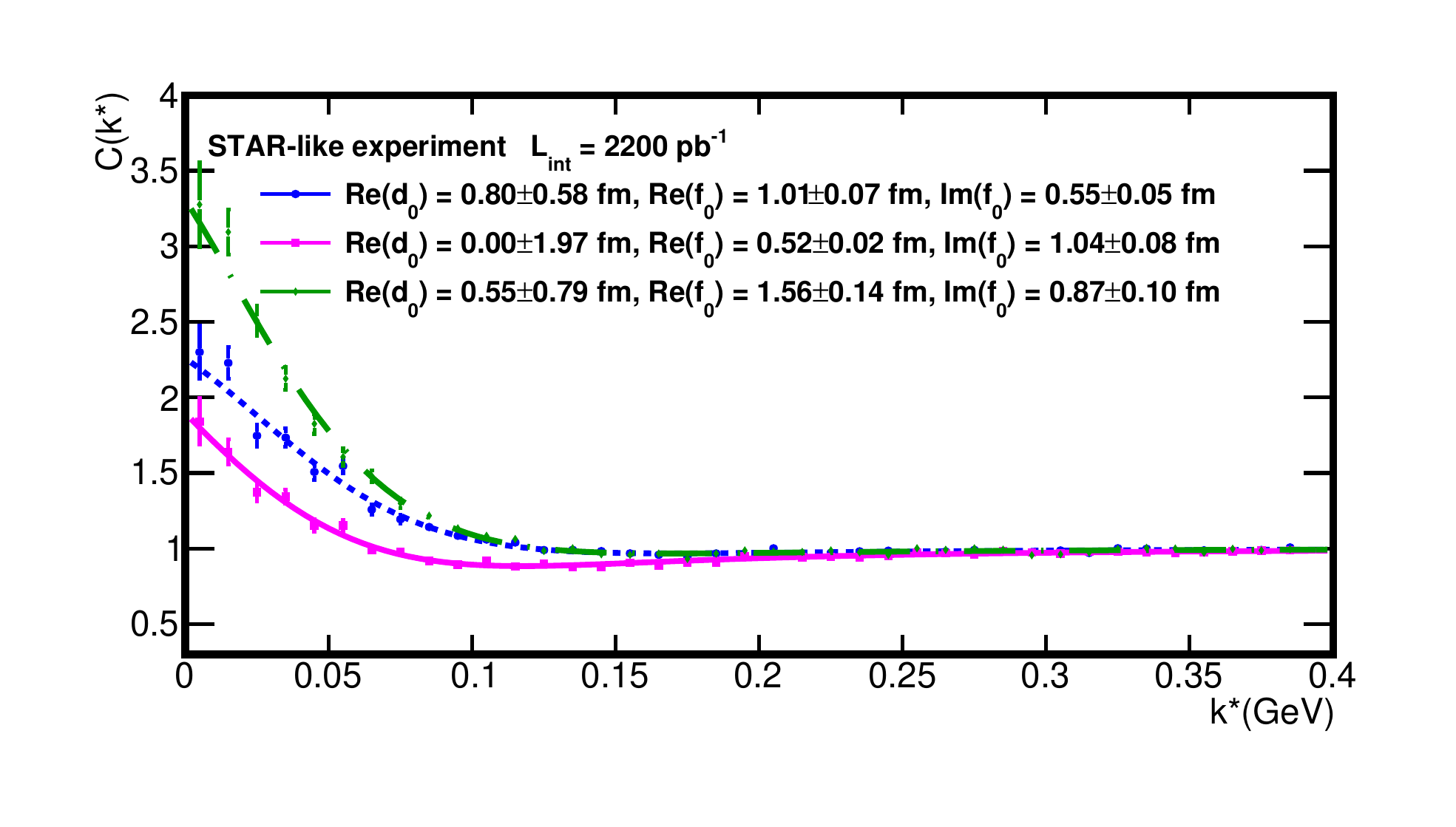}}

\caption{The pseudo-experimental correlation function for the STAR-like experiment  the plot (a) for the first three sets and plot (b) for the second three sets of parameters from Table~\ref{tab:input:params}, and their fit. }
\label{fig:corr:fun:STAR}       
\end{figure}

\begin{table*}
	\centering
	
	\begin{tabular}{ccccc}
		\toprule
		\multicolumn{5}{c}{LHCb-like, $\sqrt{s}=8$ TeV, $L_{\textrm int} = 2028 \, pb^{-1}$} \\
     	\midrule
		Parameter set No.& $\operatorname{Re}(d_{0}^{S})$ [fm] &$\operatorname{Re}(f_{0}^{S})$ [fm]&$\operatorname{Im}(f_{0}^{S})$[fm] & $\chi^2/{\textrm{NDF}}$\\
		\midrule
		1 & 1.00$\pm$0.215&0.20$\pm$0.001 & 0.00$\pm$0.008&29.62/36\\
		2 & 0.99$\pm$0.019&0.20$\pm$0.001&0.49$\pm$0.002& 37.17/36  \\
		3 & 1.02$\pm$0.018&0.50$\pm$0.002&0.50$\pm$0.002 & 24.23/36\\
		4 & 0.98$\pm$0.025&1.00$\pm$0.003&0.50$\pm$0.002&26.41/36 \\
		5 & 0.00$\pm$0.017&0.50$\pm$0.001&0.98$\pm$0.003&40.12/36 \\
		6 & 0.01$\pm$0.037&1.50$\pm$0.006&0.99$\pm$0.005&46.42/36\\
		\midrule
		\midrule
		\multicolumn{5}{c}{STAR-like, $\sqrt{s}=500$ GeV, $L_{\textrm int} = 400 \, pb^{-1}$} \\
		\midrule
        Parameter set No.& $\operatorname{Re}(d_{0}^{S})$ [fm] &$\operatorname{Re}(f_{0}^{S})$ [fm]&$\operatorname{Im}(f_{0}^{S})$[fm] & $\chi^2/{\textrm{NDF}}$\\

		\midrule
		1 & 0.00$\pm$1.45&0.21$\pm$0.04&0.00$\pm$0.05&37.70/36 \\
		2 &0.44$\pm$1.33&0.24$\pm$0.07&0.59$\pm$0.13& 50.53/36 \\
		3 &2.39$\pm$1.09&0.70$\pm$0.13&0.71$\pm$0.15 &43.32/36 \\
		4 & 1.38$\pm$1.15&1.07$\pm$0.14&0.47$\pm$0.11& 32.26/36 \\
		5 & 0.44$\pm$0.84& 0.54$\pm$0.10&1.30$\pm$0.24&40.89/36\\
		6 &  0.00$\pm$1.34&1.61$\pm$0.10&1.20$\pm$0.18& 34.04/36 \\
		\midrule
		\midrule
		\multicolumn{5}{c}{STAR-like, $\sqrt{s}=500$ GeV, $L_{\textrm int} = 2.2 \, fb^{-1}$} \\
		\midrule
	Parameter set No.& $\operatorname{Re}(d_{0}^{S})$ [fm] &$\operatorname{Re}(f_{0}^{S})$ [fm]&$\operatorname{Im}(f_{0}^{S})$[fm] & $\chi^2/{\textrm{NDF}}$\\		
		\midrule
		1 &1.02$\pm$5.05&0.21$\pm$0.02&0.01$\pm$ 0.07& 37.36/36\\
		2 &1.06$\pm$0.50&0.16$\pm$0.03&0.52$\pm$0.06&35.09/36 \\
		3 &1.13$\pm$0.49&0.49$\pm$0.04&0.51$\pm$0.05 &41.37/36\\
		4 &0.80$\pm$0.58&1.01$\pm$0.07&0.55$\pm$0.05&37.56/36\\
		5 &0.00$\pm$1.97&0.52$\pm$0.02&1.04$\pm$0.08&27.91/36\\
		6 &0.55$\pm$0.79&1.56$\pm$0.14&0.87$\pm$0.10&49.10/36\\
		\bottomrule
	\end{tabular}
	\caption{Interaction parameters extracted from the fit of the L-L model to femtoscopic correlation functions simulated for LHCb-like and STAR-like experiments. The parameters $r_{0}$ and $\operatorname{Im}(d_{0}^{S})$ are fixed to 1.25~fm and 0 respectively.}
	
	\label{tab:extracted:params}
\end{table*}

\begin{figure}

\subfloat[]{  \includegraphics[scale=0.80]{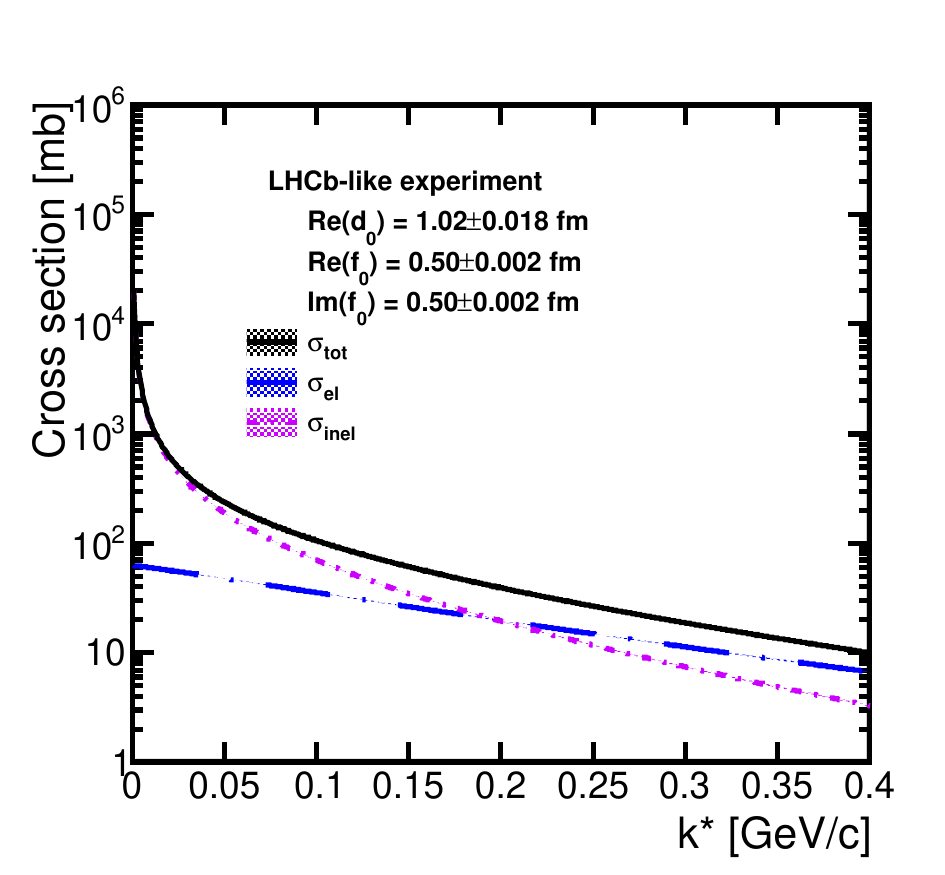}}
\vspace{-0.5cm}
\subfloat[]{  \includegraphics[scale=0.80]{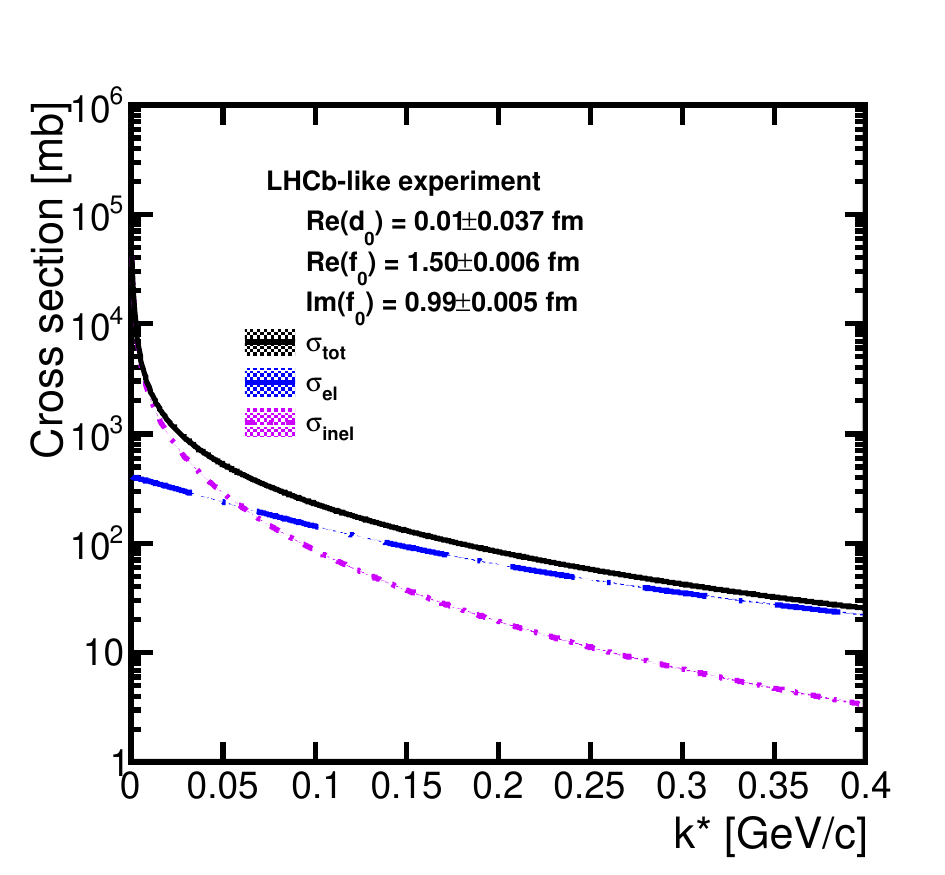}}

\caption{ Cross sections vs. $\kStar$ calculated with the parameters obtained form the fit for LHCb-like experiment the set 3 (a) and set 6 (b). The uncertainties on the plots include statistical uncertainty from the fit. }
\label{fig:res:1}       
\end{figure}

Figures \ref{fig:corr:fun:LHCb} and \ref{fig:corr:fun:STAR} show the simulated correlation functions for STAR-like and LHCb-like experiments for $L_{\textrm int} = 2.2 \, fb^{-1}$ and $L_{\textrm int} = 2.1 \, fb^{-1}$, respectively; together with fits of the L-L model. Table~\ref{tab:extracted:params} shows the scattering length and the effective range which we extracted from the fits in our feasibility studies, including the case for a STAR-like detector for \s{500}~GeV and $L_{\textrm int} = 400 \, pb^{-1}$. 

All the results in the Table \ref{tab:extracted:params} are consistent with the input values. The precision registered for STAR-like detector and the data collected in 2017 ($L_{\textrm int} = 400 \, pb^{-1}$) is not sufficient for a fruitful study. However, we expect that such a measurement will yield useful results for data taking campaign planned at RHIC at 2023. With the expected $L_{\textrm int} = 2.2 \, fb^{-1}$, the relative uncertainties on scattering length are about $10-20\%$. In the case of the LHCb-like detector, the precision of the obtained scattering length is better than 1\% for almost all considered cases. In general, the stronger the correlation effect, the better precision of the fit results. 

We use the obtained parameters to calculate the cross section for \jpsi-hadron interactions. Fig.  \ref{fig:res:1} shows examples of elastic, inelastic, and total cross section as a function of $\kStar$. 
It demonstrates that the LHCb experiment can measure with a good precision the elastic and inelastic (break-up) \jpsi-hadron cross sections using the approach we propose and the data it has already collected.  

\section{Prospects for \jpsi-hadron femtoscopic measurements}
\label{sec:prospects}

Besides the data sets that have already been studied in this paper, other data sets with high enough \jpsi\ yield also can be considered for the study of \jpsi-hadron femtoscopic correlations. In the case of the LHCb experiment, one can add
the data recorded at \s{13} TeV  with $L_{int}$ of 1.4 $fb^{-1}$ with \jpsi ~yield of 2 million~\cite{Aaij:2017fak} to our study and enhance the statistics.
For the future runs of the LHCb experiment, the $L_{int}$ at the end of Run 3 and Run 4~\cite{lhcb-future} will be up to 23 $fb^{-1}$ and 50 $fb^{-1}$ respectively. And in the year 2030, the upgrade of the LHCb detector~\cite{LHCbCollaboration:2014tuj} will allow it to run at higher luminosity up to  300 $fb^{-1}$.
 It is worth noting that the number of quarkonium states per unit of integrated luminosity in Run 2 was higher by a factor of 5  both due to the higher cross-section at 13 TeV and to improvements in the event selection during data taking. Starting from Run 3, pp collision rate at LHCb will increase by a factor of 5 as well~\cite{Bozzi_2017,lhcb-public}. 
As for the CMS experiment, the expected  $L_{int}$ at the end of Run 3 is up to 300 $fb^{-1}$ ~\cite{Klein:2017nke}. The expected high-quality data samples should allow for more differential studies and measurement of femtoscopic correlations of \jpsi\ with identified hadrons (\jpsi-$\pi^\pm$, \jpsi-proton etc.)
For similar correlation studies for $\Upsilon$-hadron pairs, for the same data set as we used in this paper for LHCb experiment ~\cite{lhcb_1} within the  $L_{int}=2082$ $pb^{-1}$ the expected number of $\Upsilon$ is around 1.8 million. For Run 3 and Run 4, this number can be scaled to 18 and 36 million respectively, which is more than the number of \jpsi\ in STAR for future data campaign in 2023. Therefore, $\Upsilon$-hadron correlation studies will be feasible within the LHCb experiment.
Moreover, the CMS experiment is expected to register around 1.7 million of $\Upsilon$ in the LHC Run 3~\cite{osti_1647062}, providing another opportunity for $\Upsilon$-hadron femtoscopy.

\section{Conclusion}
\label{sec:con}

We proposed an experimental method to study the elastic and inelastic interaction of charmonium and bottomonium with hadrons. Quantitative understanding of these processes is important for the correct interpretation of quarkonium production measurements in heavy-ion collisions and for using quarkonium to probe the properties of hot and dense nuclear matter. The proposed approach is straightforward and the experiments employed a similar strategy to study final-state interactions with success.

The method uses the femtoscopic correlation function and the Lednicky-Lyuboshitz analytical model to extract the scattering length and the effective range of the quarkonium-hadron interaction at low relative momentum. We demonstrated that such a measurement is already feasible at the LHCb experiment with $pp$ data at \s{8}~TeV collected in 2012, and it is within a reach of the STAR experiment in 2023, when a data sample of $pp$ collisions at \s{510}~GeV with luminosity of 2.2 $fb^{-1}$ is planned. 

Our feasibility study showed that LHCb can already measure both elastic and inelastic (break-up) cross sections in a hadronic matter as a function of relative momentum $\kStar$.
Such a study is more viable for LHCb future data taking campaigns when the expected luminosity will be enhanced.

\begin{acknowledgements}
We thank Adam Kisiel for providing a numerical code for computing correlation functions using the Lednicky and Lyuboshitz analytical model. The work was supported by the National Science Centre, Poland, under the research project  ``Study of quark-gluon matter properties using heavy-quark correlations'' no. 2018/30/E/ST2/00089. The Work of Leszek Kosarzewski was supported by the project LTT18002 of the Ministry of Education, Youth, and Sport of the Czech Republic.

\end{acknowledgements}

%\section*{Acknowledgement}
%We thank Adam Kisiel for providing a numerical code for computing correlation functions using the Lednicky and Lyuboshitz analytical model. The work was supported by the National Science Centre, Poland, under the research project  ``Study of quark-gluon matter properties using heavy-quark correlations'', no UMO-2018/30/E/ST2/00089. Work of L.K. was supported by ...

% BibTeX users please use one of
%\bibliographystyle{spbasic}      % basic style, author-year citations
%\bibliographystyle{spmpsci}      % mathematics and physical sciences
\bibliographystyle{spphys}       % APS-like style for physics
\bibliography{HIbiblio}   % name your BibTeX data base

\end{document}